\documentclass[11pt]{article}
\usepackage{graphicx}
\usepackage{amsmath}
\usepackage{amssymb}
\usepackage{amsxtra}

\newcommand{\ba}{\begin{eqnarray}}
\newcommand{\ea}{\end{eqnarray}}
\newcommand{\be}{\begin{equation}}
\newcommand{\ee}{\end{equation}}
\title{Describing 2-TeV scale $W_L W_L$ resonances with Unitarized Effective Theory}
\author{Felipe J. Llanes-Estrada~\footnote{Speaker, {\tt fllanes@fis.ucm.es}}, Antonio Dobado and Rafael L. Delgado,\\
Dept. Fisica Teorica I, Univ. Complutense, 28040 Madrid, Spain.}

\begin{document}
\maketitle

\begin{abstract}
The LHC is now exploring the 1-3 TeV scale where resonances of the Electroweak Symmetry Breaking Sector might exist. If so, Unitarized Effective Theory can be used to describe the data with all the constraints of unitarity, causality and global-symmetry breaking, and to 
find the resonance positions in the complex $s$-plane.
From any resonances found, one can infer the parameters of the
universal Effective Lagrangian, and those may be used to inform higher-energy theories (UV completions) that can be matched to it. 
We exemplify with two-body resonances in the coupled channels  $hh$ and $W_LW_L- Z_LZ_L$ employing the Equivalence Theorem and comment on the apparent excess in the ATLAS dijet data at 2 TeV.
\end{abstract}

\section{Non-linear EFT for $W_LW_L$ and $hh$ 
}
\label{sec:Lag}

The LHC has found a scalar boson with $m_h=125$ GeV and not much more. It is natural to describe the Electroweak Symmetry Breaking Sector of the Standard Model (SM) in terms of the low-energy spectrum alone. The resulting effective Lagrangian for the Higgs-like particle $h$ and the longitudinal gauge bosons $W_L,\ Z_L \sim \omega^a$ in the non-linear representation appropriate for the global symmetry breaking scheme 
$SU(2)\times SU(2) \to SU(2)_c$ (leaving the approximate custodial subgroup as a good isospin symmetry) is as given by us~\cite{Delgado:2013loa}, the Barcelona group ~\cite{Espriu} and others~\cite{Alonso:2014wta,Kilian:2014zja,Buchalla:2015wfa},
\ba \label{bosonLagrangian} {\cal L}
& = & \frac{1}{2}\left[1 +2 a \frac{h}{v} +b\left(\frac{h}{v}\right)^2\right]
\partial_\mu \omega^i \partial^\mu
\omega^j\left(\delta_{ij}+\frac{\omega^i\omega^j}{v^2}\right) \nonumber
+\frac{1}{2}\partial_\mu h \partial^\mu h \nonumber  \\
 & + & \frac{4 a_4}{v^4}\partial_\mu \omega^i\partial_\nu \omega^i\partial^\mu
 \omega^j\partial^\nu \omega^j +
\frac{4 a_5}{v^4}\partial_\mu \omega^i\partial^\mu \omega^i\partial_\nu
\omega^j\partial^\nu \omega^j  +\frac{g}{v^4} (\partial_\mu h \partial^\mu h )^2
 \nonumber   \\
 & + & \frac{2 d}{v^4} \partial_\mu h\partial^\mu h\partial_\nu \omega^i
 \partial^\nu\omega^i
+\frac{2 e}{v^4} \partial_\mu h\partial^\nu h\partial^\mu \omega^i
\partial_\nu\omega^i
\ea

The parameters of this Lagrangian, neglecting the masses of all quasi-Goldstone bosons $\omega^a$ and of the Higgs $h$, adequate to explore the energy region 1-3 TeV $\gg$ 100 GeV, are seven. Their status is given in table~\ref{tab:params}.

\begin{table}[h]
\begin{center}
\begin{tabular}{|cc|ccccc|}
\hline
$a$ & $b$ & $a_4$ & $a_5$ & $g$ & $d$ & $e$ \\ \hline
$(0.88,1.34)$ & $\in  (-1,3)a^2$ (this work) & 0? & 0? & 0? & 0? & 0? \\
\hline
\end{tabular}
\caption{\label{tab:params} From the ATLAS and CMS reported~\cite{ATLAS:2014yka} $hWW$, $hZZ$ couplings   one can infer the approximate constraint shown on $a$ at 2$\sigma$ level (a recent communication to the LHCP2015 conf. finds similar results~\cite{pressrelease}).
In our recent 
work~\cite{Delgado:2015kxa,Delgado:2015kxa2} on unitarized perturbation theory we could also put a coarse constraint on $b$ due to the absence of a coupled-channel resonance in $hh-\omega\omega$ (the second channel is visible while the first is much harder). Basically no bounds have been reported on the NLO parameters: their SM value is zero.}
\end{center}
\end{table}
We emphasize that with seven parameters, this is a reasonably manageable Lagrangian for LHC exploration of electroweak symmetry breaking, granted, under the approximation of $M_W\simeq M_Z \simeq m_h \simeq 0$ which is fair enough in the TeV region, and this is in contrast to the very large parameter space of the fully fledged effective theory~\cite{Alonso:2014wta}.

The perturbative scattering amplitudes 
$A_{I}^J(s)= A^{(LO)}_{IJ}(s)+A^{(NLO)}_{IJ}(s)\dots$   
for $\omega\omega$ and $hh$, projected into partial waves, are given to NLO in~\cite{Delgado:2015kxa}. For example, the LO 
amplitudes of $I=$ 0, 1 and 2, proportional to $(1-a^2)$, and the channel-coupling amplitude $\omega\omega\to hh$, to $(a^2-b)$,
\begin{eqnarray} \label{LOamps}
A_0^0(s) & = & \frac{1}{16 \pi v^2} (1-a^2) s  \nonumber \\
A_1^1(s) & = & \frac{1}{96 \pi v^2} (1-a^2) s  \nonumber \\
A_2^0(s) & = & -\frac{1}{32 \pi v^2} (1-a^2)s \nonumber \\
M^0 (s)  & = & \frac{\sqrt{3}}{32 \pi v^2} (a^2-b)s \nonumber 
\end{eqnarray}
 show how a tiny separation of the parameters from the SM value leads to an energy-growing, eventually strongly interacting set of amplitudes.

Including the NLO, these amplitudes take a form characteristic of chiral perturbation theory
\be \label{pertamplitude}
A_{IJ}^{(LO+NLO)}(s) = K s + \left( B(\mu)+D\log\frac{s}{\mu^2}+E\log\frac{-s}{\mu^2}\right) s^2 
\ee
with a left cut carried by the $Ds^2\log s$ term, a right cut in the $Es^2\log (-s)$ term, and the $Ks+Bs^2$ tree-level polynomial. $B$, $D$ and $E$ have been calculated, reported in~\cite{Delgado:2015kxa} and allow for perturbative renormalizability, where the  chiral counterterms contained in $B$ absorb one-loop divergences from iterating the tree-level Lagrangian and run  to make Eq.~(\ref{pertamplitude}) scale invariant.

The energy reach of the Effective Theory with the Lagrangian density in 
Eq.~(\ref{bosonLagrangian}) is nominally $4\pi v\sim 3$ TeV.
If the LHC finds no clear new phenomenon through this scale,  experimental data on $W_LW_L$ spectra can eventually be compared with the effective theory predictions. In this precision work, separations of $a$ from 1 or of $b$ from $a^2$ or any NLO parameter from 0 can then be used to predict the scale of new physics, or if measurements are null, at least to constrain it.

\section{Resonances}
On the other hand, if the LHC finds new resonances that couple to two longitudinal gauge bosons (and potentially also to two Higgs bosons), then a purely perturbative approach is inadequate. A deffect of the amplitudes 
in Eq.~(\ref{pertamplitude}) is that they violate the unitarity relation
${\rm Im} A_{IJ} = |A_{IJ}|^2 $, which is satisfied only order by order in perturbation theory, namely ${\rm Im} A^{({\rm NLO})}_{IJ} = 
| A^{({\rm LO})}_{IJ}|^2$. 
This introduces an error which is only acceptably small when $s$ is much smaller than the mass of the first resonance in the $IJ$ channel. But of course, since near resonances the imaginary part of the amplitude is large, the effective theory is of no use there. The solution is sometimes called Unitarized Effective Theory and is described in subsection~(\ref{subsec:unitarization}).

\subsection{Unitarization} \label{subsec:unitarization}
Unitarization of effective theory amplitudes is a technique well-known~\cite{Lehmann:1972kv} in hadron physics that we describe only briefly.
It is possible because scattering amplitudes in field theory are very constrained functions due to Lorentz invariance, causality and unitarity. Dispersion relations, known from old in optics, are a way of incorporating all the constraints~\cite{Mandelstam:1958xc}  leaving little freedom to determine the amplitudes, though they remain ambiguous without dynamical knowledge. To fully obtain them though, one needs a few key numbers which are provided by the effective theory at low-energy (see the lectures~\cite{Truong:1990du} for an introduction). 
This powerful method of combining dispersion relations with effective theory, which basically exhausts all underlying-model independent information in the experimental data for two-body channels, was deployed for the electroweak symmetry breaking sector early on~\cite{Dobado:1989gr}.  Usually the resulting amplitudes for $W_LW_L\sim \omega\omega$ scattering are encoded in simple algebraic forms that avoid the complications of the dispersion relations, such as the K-matrix~\cite{Kilian:2014zja}  that introduce a small amount of model dependence in the discussion. 

To address this, we have compared~\cite{Delgado:2015kxa} three unitarization methods that agree in predicting the same resonances at the same positions within 1 to 10\% when all three can be used. 
These are the Inverse Amplitude Method, the N/D method, and an improved version of the K-matrix method that ensures complex-plane analyticity where appropriate. Table~\ref{tab:unitmethods} shows the $IJ$ channels where each one is currently applicable in the Electroweak sector.
\begin{table}
\caption{\label{tab:unitmethods} Channels where each unitarization method can currently be used.}
\begin{center}
\begin{tabular}{|c|*{5}{c}|}
\hline
$IJ$    &   00  & 02    & 11    & 20    & 22    \\
\hline
Method  & All  & N/D, IK&IAM    & All & N/D, IK \\
\hline
\end{tabular}
\end{center}
\end{table}

\begin{figure}
\includegraphics[width=0.48\textwidth]{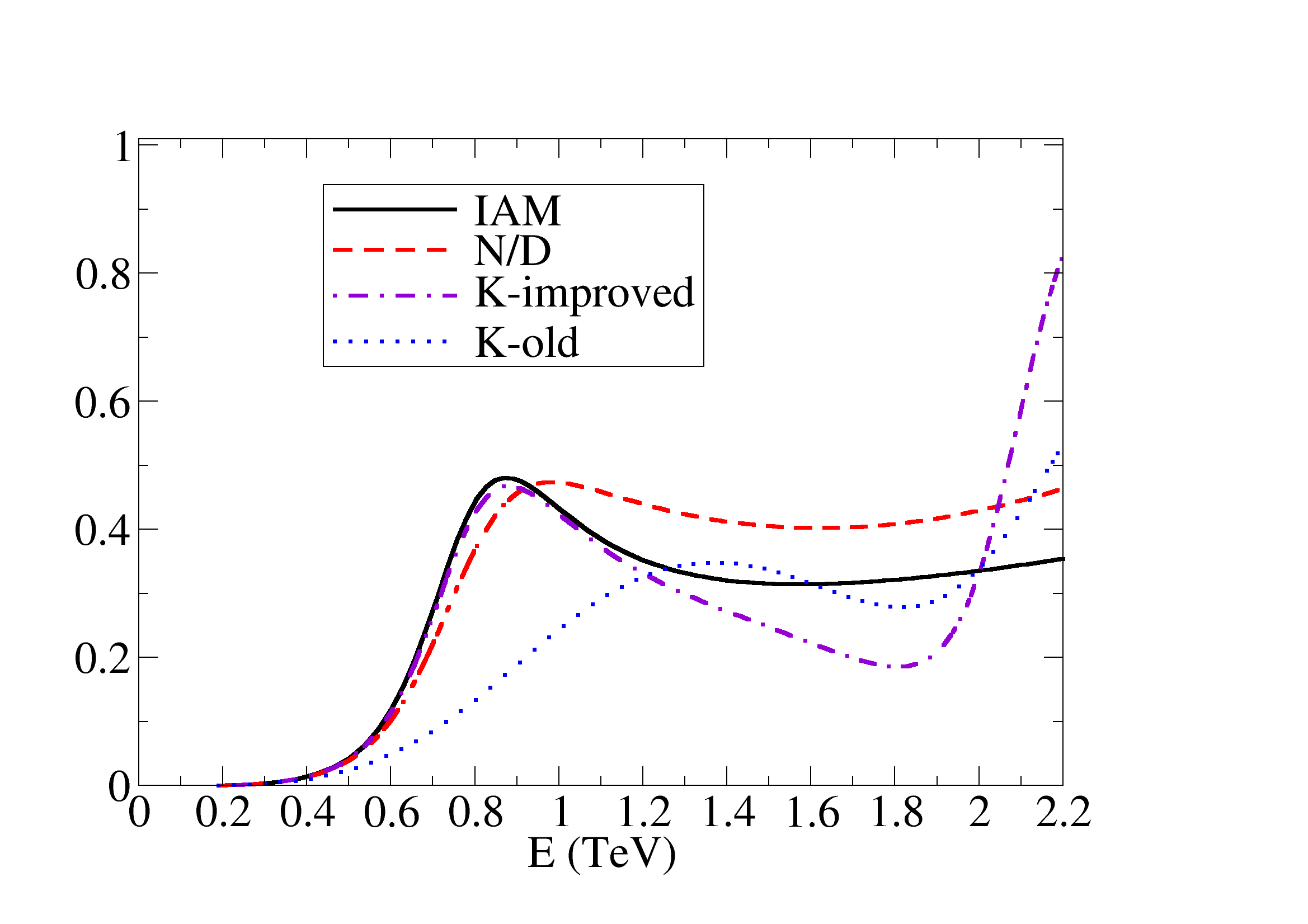}
\includegraphics[width=0.48\textwidth]{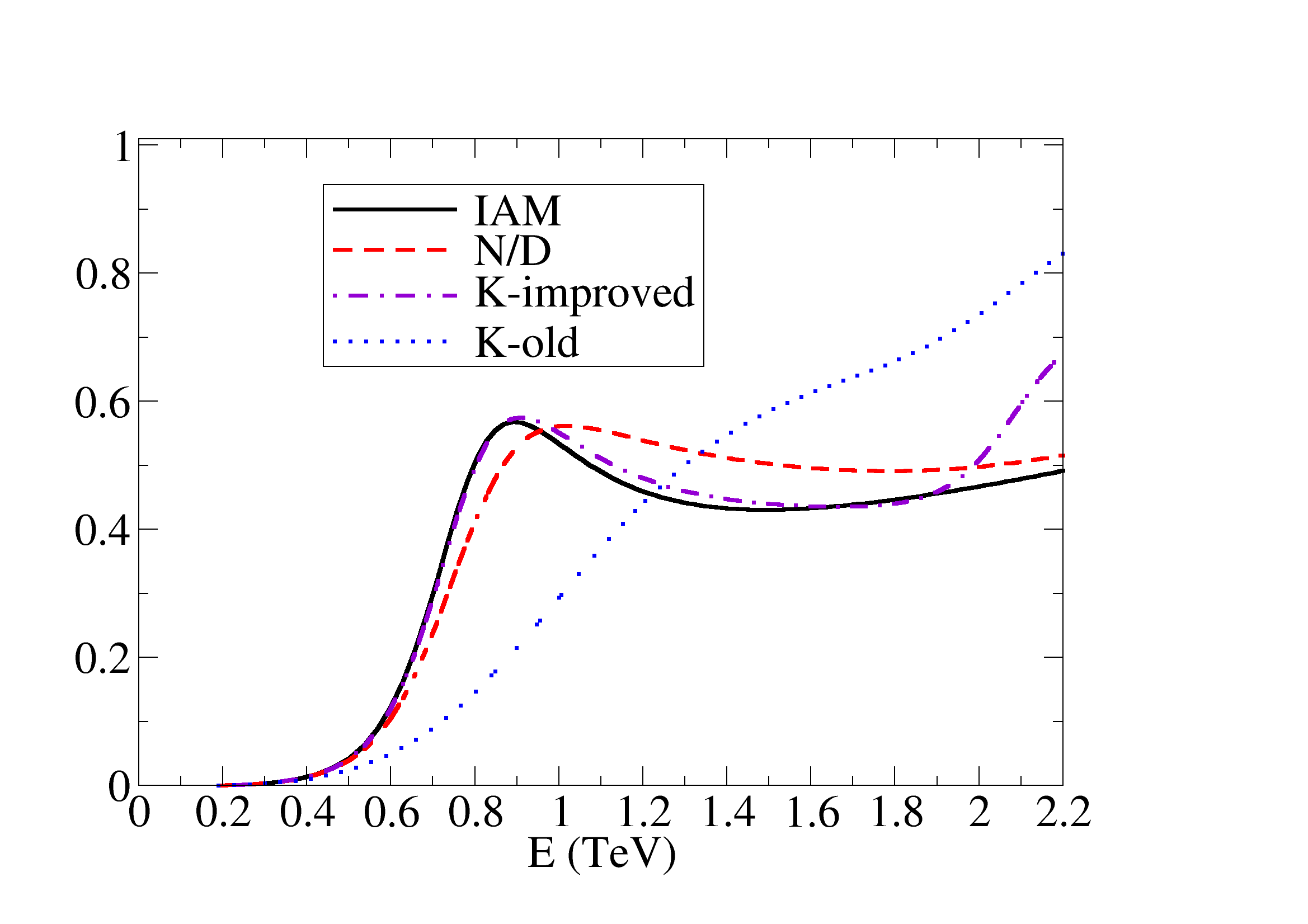} \\
\begin{minipage}{0.48\textwidth}
\includegraphics[width=0.95\textwidth]{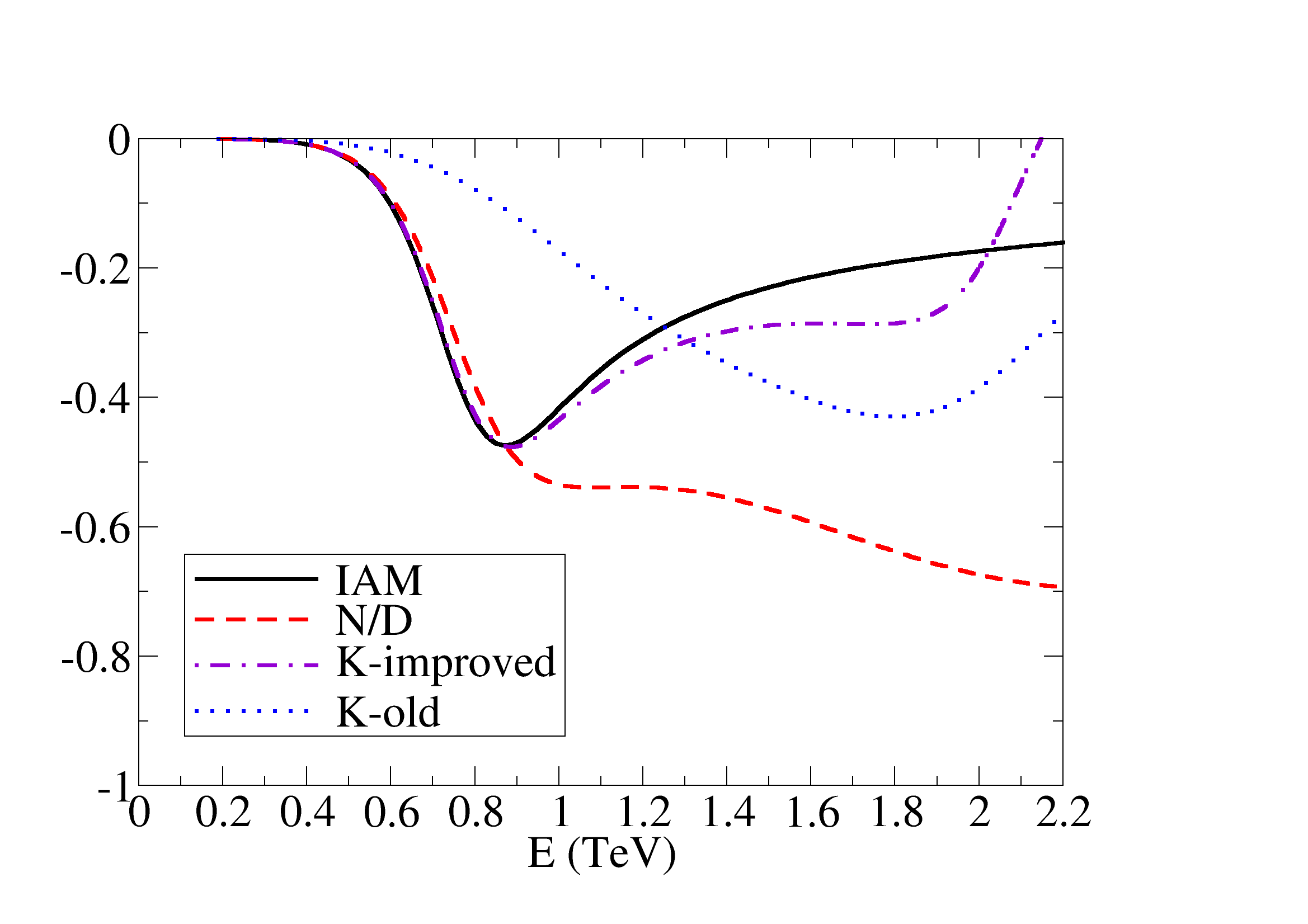}
\end{minipage}
\begin{minipage}{0.48\textwidth}
\caption{\label{fig:IAM} Comparison of three unitarization methods for the imaginary parts of the $IJ=00$ amplitudes. Clockwise from top left, $\omega\omega$, $hh$ and channel-coupling $\omega\omega\to hh$ (parameters in the text). A scalar resonance is visible in all, and the unitarization methods with correct analytic properties closely agree.}
\end{minipage}
\end{figure}

As an example, consider the Inverse Amplitude Method. In its simplest form it requires two orders of the perturbative expansion, that are combined in the following simple formula, 
\begin{equation} \label{IAM}
A_{IJ} = \frac{\left( A^{(LO)}_{IJ}\right)^2}{A^{(LO)}_{IJ}-A^{(NLO)}_{IJ}}\ .
\end{equation}
 To obtain it, one realizes that a dispersion relation for $A(s)$ may be exact but of little use because of insufficient low-energy information. On the contrary, a dispersion relation for the perturbative $A^{(LO)}+A^{(NLO)}$ can be fully studied, but it is trivial because the perturbative amplitude is known everywhere.  The trick is to write one for $\left( A^{(LO)}\right)^2 A^{-1}$ (hence the name ``Inverse Amplitude Method'') because the integral over the right, unitarity cut of $1/A$ is exactly calculable when only two-body channels are important. The result is the formula in Eq.~(\ref{IAM}). Its generalization to two (massless) channels is straightforward by turning the quantities therein into matrices, each element being an elastic $\omega\omega\to \omega\omega$, $hh\to hh$ or a cross-channel $\omega\omega\to hh$ amplitude. In figure~\ref{fig:IAM} we show the IAM and also the other two methods with NLO parameters set to 0 at a scale of $\mu=3$ TeV and with LO parameters $a=0.88$ and $b=3$. This set generates a characteristic coupled-channel resonance seen in all three amplitudes. 

The variable $s$ in Eq.~(\ref{IAM}) may be extended to the complex plane, allowing to search for resonances in its second Riemann sheet. We locate the pole positions and report selected ones below in subsection~\ref{subsec:polessecond}.

\subsection{ATLAS excess in two-jet events}
\begin{figure}
\centerline{\includegraphics[width=0.475\textwidth]{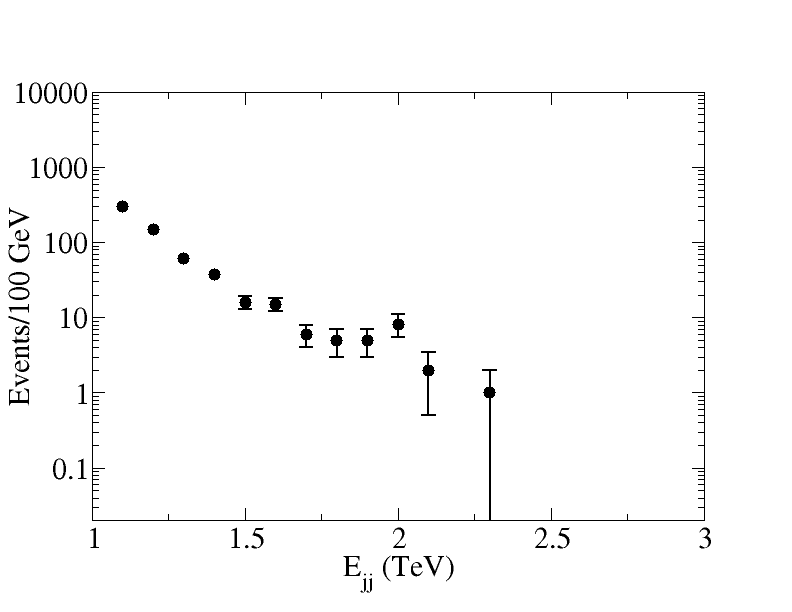}
\includegraphics[width=0.475\textwidth]{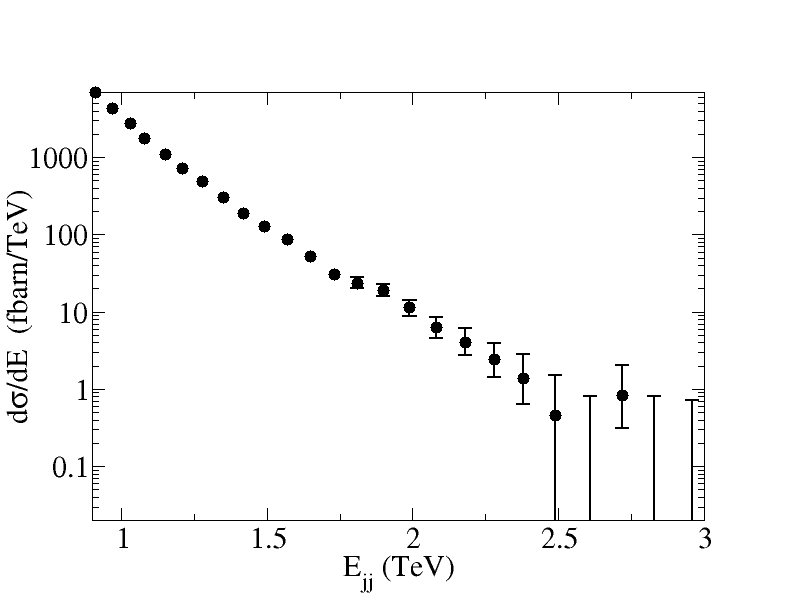}
}
\caption{\label{fig:ATLASdata} Left: rerendering of the ATLAS data\cite{Aad:2015owa} for $WZ\to 2\ {\rm jet}$ in pp collisions at the LHC, that shows a slight excess at 2 TeV (same in the other isospin combinations $WW$ and $ZZ$, not shown).
Right: CMS data~\cite{Khachatryan:2014hpa} in the same 2-jet channel with jets tagged as vector bosons. Here the collaboration provides the absolute normalization of the cross-section. No excess is visible at 2 TeV (if at all, a tiny one at 1.8-1.9 TeV).}
\end{figure}

The interest in TeV-scale resonances has recently rekindled because of an apparent excess in ATLAS data~\cite{Aad:2015owa} plotted in figure~\ref{fig:ATLASdata} together with comparable, older CMS data~\cite{Khachatryan:2014hpa} that does not show such an enhancement.

The excess is seen in two-jet events, each one containing the entire debris  of a respective gauge boson. Their invariant mass reconstruction allows the assignment of a $W$ or of a $Z$ tag (82 and 91 GeV respectively) but the experimental error makes the identification loose, so that the three-channels cross-feed and we should not take seriously the excess to be seen in all three yet. 
Because $WZ$ is a charged channel, an $I=0$ resonance cannot decay there. Likewise $ZZ$ cannot come from an $I=1$ resonance because the corresponding Clebsch-Gordan coefficient $\langle 1 0 1 0 | 1 0 \rangle$ vanishes. 
A combination of both isoscalar and isovector could explain all three signals simultaneously, as would also an isotensor $I=2$ resonance. In the isotensor case, the resonance should be visible in the doubly charged channel $W^+ W^+$ whereas not in the other (to tag the charge requires to study leptonic decays instead of jets, so it is a whole other measurement, but worth carrying out).

Numerous models have  been proposed to explain the presumed excess, but the model-independent information is still sparse~\cite{Allanach:2015hba}. 

One statement that we can make, based on the so-called KSFR relation that the IAM naturally incorporates (as do broad classes of theories such as Composite Higgs models~\cite{Kaplan:1983sm} with vector resonances~\cite{Barducci:2015oza}), is that if a $\rho$-like isovector resonance is in the ATLAS data, it will be quite narrower than the bump seen (perhaps broadened due to experimental resolution).
The relation, given here in the absence of further channels~\cite{Delgado:2015kxa2}, links the mass and width of the isovector resonance with the low-energy constants $v$ and $a$ in a quite striking manner,
\begin{equation} \label{KSFR}
\Gamma^{\rm IAM} = \frac{M^3_{\rm IAM}}{96\pi v^2}(1-a^2)\ .
\end{equation}
For $M\sim 2$ TeV and $\Gamma\sim 0.2$ TeV as obtained by rule of thumb in fig.~\ref{fig:ATLASdata}, one gets $a\sim 0.73$ which is in tension with the ATLAS-deduced bound $a \arrowvert_{2\sigma}>0.88$ at 4-5$\sigma$ level; Eq.~(\ref{KSFR}) predicts that an isovector $W_LW_L$ resonance at 2 TeV, with present understanding of the low-energy constants, needs to have a width of order 50 GeV at most.

\subsection{IAM parameter map} \label{subsec:polessecond}
At last, we map out part of the seven-parameter space in search for resonances at 2 TeV that can be brought to bear on the new ATLAS data.

\begin{figure}
\centerline{
\includegraphics[width=0.48\textwidth]{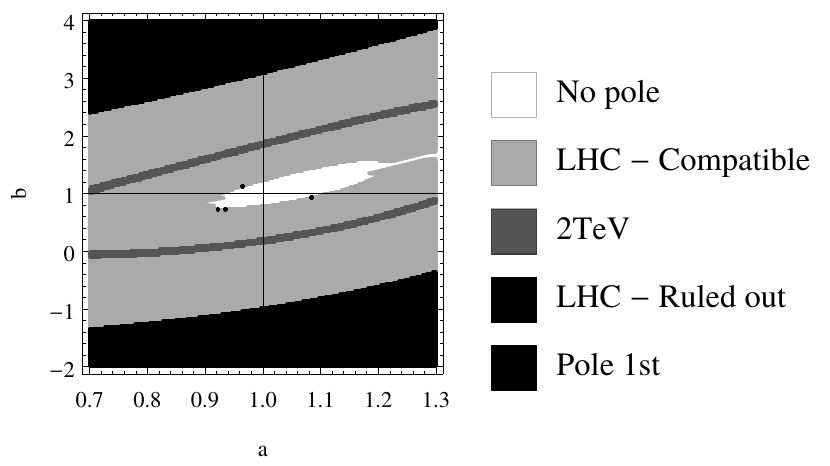}
\includegraphics[width=0.48\textwidth]{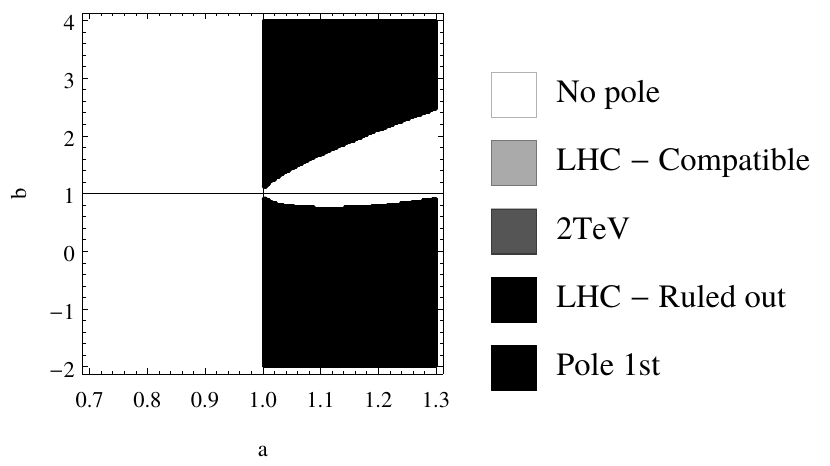}    }
\begin{minipage}{0.48\textwidth}
\includegraphics[width=0.95\textwidth]{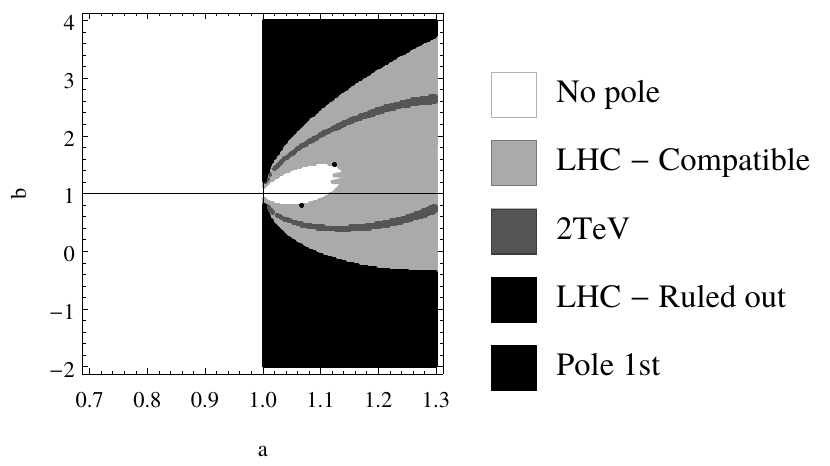}
\end{minipage}
\begin{minipage}{0.48\textwidth}
\caption{\label{fig:isoall} 
 We explore the $ab$ parameter space in search for resonant poles of $\omega\omega$ scattering; clockwise from top left, $IJ=00$, $11$ and $20$. 
}
\end{minipage}
\end{figure}

For $a<1$ the scalar-isoscalar channel can be resonant from the LO Lagrangian alone (generating a $\sigma$-like resonance that was described in
~\cite{Delgado:2013loa}). In fact, even for $a=1$, there is a resonance generated for large enough $b$ that oscillates between the $\omega\omega$ and $hh$, a ``pinball'' resonance, reported in~\cite{Delgado:2015kxa}.
This can be seen in the left plot of figure~\ref{fig:isoall}, where, for $a<1$ so that $(1-a^2)>1$ there is a pole in the second Riemann sheet. 

The isoscalar wave  resonates for a broad swipe of $ab$ parameter space, and near 2 TeV (the thin band), though the structure is generally broad, and feeds the $WW$ and $ZZ$ channels seen in the ATLAS data. In that case, the charged $WZ$ experimental excess must be ascribed to misidentification of one of the two bosons, since an isoscalar resonance is of necessity neutral.

For $a>1$  an isotensor resonance exists (see again fig.~\ref{fig:isoall}, bottom plot). This is possible for $a>1$ (light gray band marked "LHC compatible") as the LO amplitude in Eq.~(\ref{LOamps}) becomes attractive.  Of course, for this negative sign of $(1-a^2)$, as seen in Eq.~(\ref{LOamps}), the usual roles of the isoscalar and isotensor waves are reversed, with the first now being repulsive. 

In a narrow curved strip (middle gray, immersed in that band) this resonance appears at about 2 TeV and can decay to all of $WW$, $WZ$ and $ZZ$ charge-channels. The darkest area  corresponds here to ``LHC ruled out'' and means that the resonance is light and might already be excluded.

We need to make sure that the other waves don't present causality-violating poles in the first Riemann sheet that rule out a certain parameter region. Returning to figure~\ref{fig:isoall} we see that the isovector wave indeed violates causality for much of the parameter space where the isotensor resonance exists, though there are perhaps small patches where the isotensor resonance is still allowed, for not too large values of $b$.

Since this allowed parameter space is so small and because, even if the isotensor resonance were there its production cross-section would be smaller (requiring two intermediate $W$ bosons) than the production of an isovector one as reported in~\cite{Dobado:2015hha}, we proceed to the NLO amplitude.

\begin{figure}
\includegraphics[width=0.48\textwidth]{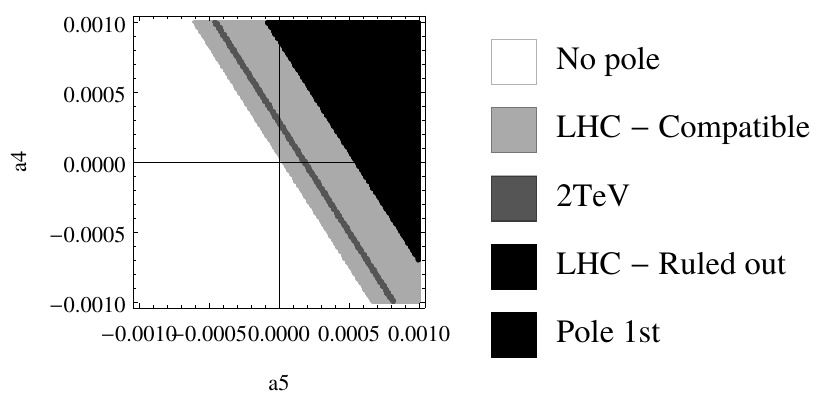}
\includegraphics[width=0.48\textwidth]{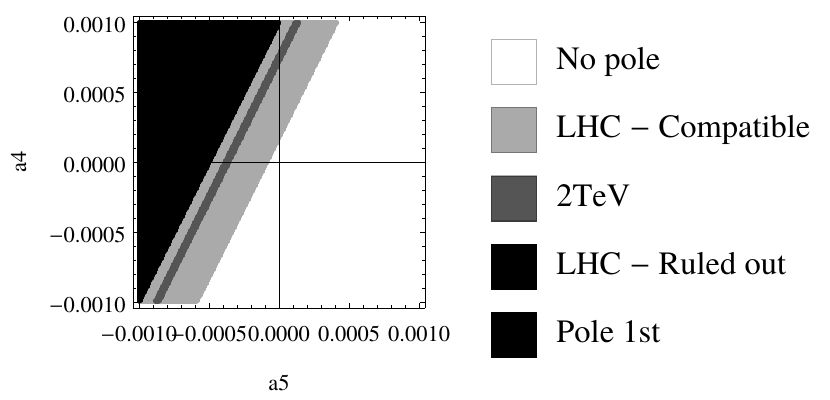}\\
\begin{minipage}{0.48\textwidth}
\includegraphics[width=0.95\textwidth]{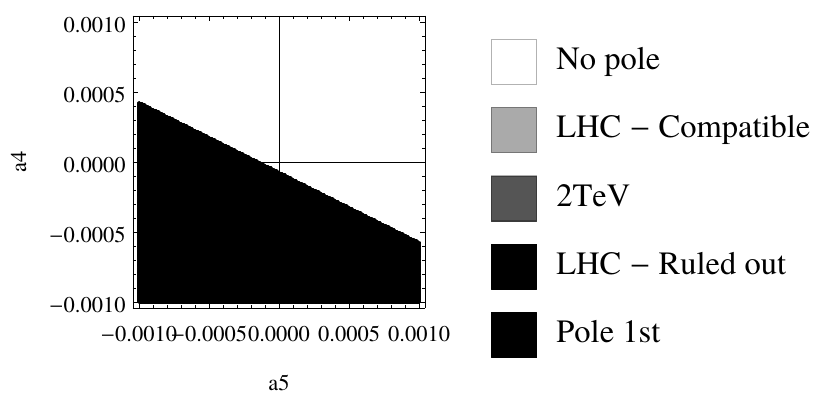}
\end{minipage}
\begin{minipage}{0.48\textwidth}
\caption{\label{fig:allthree}
Searches for complex-plane poles as function of the NLO parameters $a_4$ and $a_5$ for fixed $a=0.95$ and $b=a^2$.
Clockwise from left top, $IJ=00$, $11$, $20$.}
\end{minipage}
\end{figure}

We likewise look for poles in the complex $s$ plane
as function of the $a_4$, $a_5$ parameters with fixed $a=0.95$ and $b=1$, as shown in fig.~\ref{fig:allthree}.
The bottom plot shows how a large swath of parameter space towards negative $a_4$ is excluded by displaying a pole in the first Riemann sheet of the $20$ channel. Because here we chose $a<1$, this channel does not resonate in the second sheet, whereas the scalar one (left, top plot) does, as well as the $11$ channel (that is seen, by comparing with fig.~\ref{fig:isoall}, to present ``intrinsic'' resonances driven by the NLO counterterms).

The two diagonal bands in the $00$ and $11$ channels that support poles at around 2 TeV intersect for slightly negative $a_5$ and $a_4$ of order $5\times 10^{-4}$. There, we find both isoscalar and isovector poles, that jointly could explain all of the extant $WW$, $WZ$ and $ZZ$ excesses in two-jet data.

\section{Conclusion}

\begin{figure}[tbh]
\centerline{\includegraphics[width=0.6\textwidth]{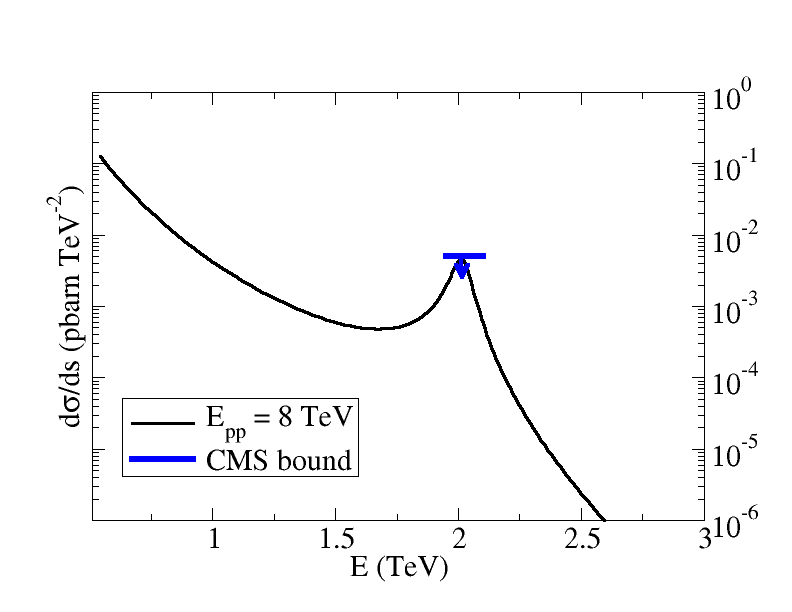}}
\caption{\label{fig:withCMS}  Tree-level $W$ production of $\omega\omega$ from~\cite{Dobado:2015hha} in the presence of resonant final-state
interactions with parameters $a=0.9$, $b=a^2$, $a_4=7\times 10^{-4}$ (at $\mu=3$ TeV). Also shown is the CMS upper bound on the cross-section obtained from fig~\ref{fig:ATLASdata}.} 
\end{figure}

The LHC is now taking data at 13 TeV and production cross-sections sizeably increase. This is necessary as the typical $\sigma$ for $\omega\omega$ resonances are currently at or below the LHC sensitivity limit as shown in fig.~\ref{fig:withCMS}. The large rate at which a resonance would have to be produced to explain the ATLAS excess is a bit puzzling.

We hope that this ATLAS excess will soon be confirmed or refuted. In any case, the combination of effective theory and unitarity, as encoded for example in the IAM, is a powerful tool to describe data up to 3 TeV of energy  in the electroweak sector if new, strongly interacting phenomena appear, with only few independent parameters. The content of new, Beyond the Standard Model theories, can then be matched onto those parameters for quick tests of their phenomenological viability.

\section*{Acknowledgements}
We owe many discussions with  J. J. Sanz Cillero and  D. Espriu. 
FLE thanks the organizers of the Bled workshop ``What comes beyond the SM" for their hospitality and encouragement.
Work partially supported by Spanish Excellence Network on Hadronic Physics FIS2014-57026-REDT,
and grants UCM:910309, MINECO:FPA2014-53375-C2-1-P, BES-2012-056054 (RLD).


\end{document}